\documentclass[doublecol]{epl2}

\usepackage{amsmath,amssymb,bm}
\usepackage{graphicx,color}

\newcommand{\hide}[1]{{}}
\newcommand{\delete}[1]{\textcolor{red}{}}
\newcommand{\gammadot}{\ensuremath{\dot{\gamma}}}
\newcommand{\Pe}{\ensuremath{\mathrm{Pe^{\ast}}}}

\newcommand{\Uwall}{\ensuremath{U_{\mathrm{wall}}}}
\newcommand{\drna}{\ensuremath{d\bm{r}^{\mathrm{na}}}}

\title{Shear-induced anisotropic decay of correlations in hard-sphere colloidal glasses}

\author{V. Chikkadi\inst{1} \and S. Mandal\inst{2} \and B. Nienhuis\inst{1} \and D. Raabe\inst{2} \and F. Varnik\inst{2,3} \and P. Schall\inst{1}}
\shortauthor{V. Chikkadi \etal}

\institute{
  \inst{1} Institute of Physics, University of Amsterdam, Science Park 904, 1098 XH Amsterdam, The Netherlands,\\
  \inst{2} Max-Planck-Institut f\"ur Eisenforschung, 40237 D\"usseldorf, Germany,\\
  \inst{3} Interdisciplinary Centre for Advanced Materials Simulation (ICAMS), Ruhr-University Bochum, Germany
}


\abstract{
Spatial correlations of microscopic fluctuations are investigated via real-space experiments and computer simulations of colloidal glasses under steady shear. It is shown that while the distribution of one-particle fluctuations is always isotropic regardless of the relative importance of shear as compared to thermal fluctuations, their spatial correlations show a marked sensitivity to the competition between shear-induced and thermally activated relaxation. Correlations are isotropic in the thermally dominated regime, but develop strong anisotropy as shear dominates the dynamics of microscopic fluctuations. We discuss the relevance of this observation for a better understanding of flow heterogeneity in sheared amorphous solids.}

\begin{document}

\maketitle

\section{Introduction}
The nature of microscopic fluctuations in the relaxation and flow of glasses is a topic of great current interest because of their importance to a wide range of materials including foams, emulsions, granulates, and colloidal and molecular glasses. Thermal glasses such as molecular and colloidal glasses exhibit structural relaxation due to thermally induced rearrangements \cite{GlassesRelaxation,weeks2000,schall2007}, while athermal amorphous materials such as granulates and foams relax typically by external forces such as applied shear that drive their flow \cite{athermalRelaxation,Schall_VanHecke2010}. It has been suggested that the applied shear acts as an effective temperature that drives microscopic fluctuations in an isotropic way \cite{FDT_grain,FDT_barrat,haxton_liu07}. Such a concept is attractive since it opens a possibility of the generalized fluctuation dissipation relation under a nonequilibrium situation. This idea is also motivated by the jamming phase diagram \cite{liu_nagel98} that depicts applied shear stress as an alternative way to fluidize the glass, in an attempt to unify amorphous materials in a common framework. Supported by simulations of sheared glasses \cite{FDT_barrat,haxton_liu07}, the current models of plasticity of amorphous materials \cite{falk_langer98,langerSTZ} assume that the rate of activated events is controlled by an isotropic effective temperature that is dependent on the shear rate. While these models have been quite successful in describing many aspects regarding the plastic response of amorphous solids to external load such as simple shear, many other issues such as flow heterogeneity and shear banding still remain under debate \cite{VarnikPRL_03, BesselingPRL_10, MandalPRL_12}. Some additional ingredient seems, therefore, to be necessary for a full description of the deformation behavior of amorphous solids.

Here, we focus on the issue of anisotropy. Since the applied shear imposes a directionality of motion, this should reflect itself in the nature of microscopic fluctuations and/or their correlations. In fact, some evidence of direction-dependent correlations comes from quasistatic computer simulations of two-dimensional Lennard-Jones glasses at zero \cite{maloney_robbins09, Lemaitre_Caroli} and finite \cite{furukawa_tanaka09} temperatures. Direct observation of these fluctuations in molecular glasses, however, is prohibited by the small molecular length scales. We use real-space experiments and molecular dynamics simulations of colloidal hard spheres to investigate the effect of shear on the microscopic fluctuations. Hard-sphere colloidal glasses provide good model systems for molecular glasses, with the advantage that the motion of the individual particles can be followed directly in real-space and time using confocal microscopy. The direct imaging of displacement fluctuations allows us to identify a novel kind of correlation with stress-dependent anisotropic scaling. We determine correlations by following fluctuations of non-affine displacements directly in real space, and we investigate their direction-dependence. By varying the applied shear rate, we probe two different relaxation regimes: the slowly driven "viscous" regime where the shear rate $\dot{\gamma}$ is smaller than the inverse structural relaxation time, $\tau$, so that relaxation occurs predominantly by thermal fluctuations, and the shear-dominated regime where $\dot{\gamma}>\tau^{-1}$, so that relaxation is driven by the applied shear. In agreement with the idea of a scalar effective temperature, we find that the single particle displacement fluctuations are always isotropic, even when shear dominates the relaxation. At the same time, however, we observe a striking difference in the \textit{spatial correlations} of fluctuations: while in the slowly driven regime, correlations exhibit isotropic decay, i.e. with a direction-independent scaling exponent, in the shear-driven regime, this decay becomes anisotropic, being slowest in the flow direction. We argue that such stress-dependent anisotropic correlations can cause the ubiquitous shear banding instability of amorphous materials. These results reveal a new kind of criticality exhibited by amorphous solids on the verge of flow. The particular feature of this stress-induced criticality is the stress-dependent anisotropy of the scaling of correlations.

\section{Experiments}
Our colloidal glass consists of sterically stabilized polymethylmethacrylate (PMMA) particles suspended in a mixture of Cycloheptyl Bromide and Cis-Decalin that matches both the density and refractive index of the particles. The particles have a diameter of $\sigma = 1.3 \mu m$, with a polydisperity of $ 7~\%$. The particle volume fraction is $\phi \sim 0.6$, well within the glassy state (note that $\phi_{\text{glass}}\approx 0.58$)\cite{Williams}. The suspension is loaded in a cell between two parallel plates at a distance of $L_z = 65 \mu m$. Shear is applied by moving the top plate in the positive $x-$direction using a piezoelectric translation stage. Confocal microscopy is used to image the motion of the individual particles in a $108\mu m \times 108\mu m \times 65\mu m$ volume, and to determine their positions in three dimensions with an accuracy of $0.03\mu m$ in the horizontal, and $0.05\mu m$ in the vertical direction \cite{weeks2000}. The motion of $\sim 2 \times 10^5$ particles is then tracked during a time interval of $25~min$ by acquiring image stacks every $60s$. All measurements are taken in the steady state regime, after the glass has been sheared to a strain of $\gamma \sim 1$.

As a basic quantity, we determine non-affine displacements as follows \cite{Goldenberg2007}. For each particle, we follow nearest neighbors in time and determine the best affine tensor $\bm{\epsilon}$ that transforms the nearest neighbor vectors, $\bm{d}_i=\bm{r}_i-\bm{r}_0$, over the time interval $\delta t$. This is done by minimizing $D^2 = (1/n) {\sum_{i=1}^{n}}(\bm{d}_i(t + \delta t) - (\bm{I} + \bm{\epsilon})\cdot \bm{d}_i(t))^2$, where $\bm{I}$ is the identity matrix. $D^2$ reflects the mean-square deviation from a local affine deformation, and is an excellent  measure of local plasticity \cite{falk_langer98}. The non-affine displacements of the particles are determined both (i) via subtraction of the local flow, $\drna=\bm{r}(t+\delta t)- \bm{r}(t) - \int_0^t dt' \bm{u}(t', \bm{r}(t'))$, where $\bm{u}(t,\bm{r}(t))$ is a coarse-grained displacement field~\cite{Goldenberg2007} and (ii) as $\drna=\bm{r}(t+\delta t)-\bm{r}(t) - \bm{\epsilon} \cdot  \bm{r}(t)$. It is assumed in (ii) that the coordinate center is at rest. We have verified that these two definitions give identical results as long as the shear is homogeneous across the channel~\cite{Chikkadi2012}. In the general case of heterogeneous shear, however, the first definition is used.

To separate the effects of shear and temperature, we distinguish applied shear rates by the modified Peclet number \cite{cipelleti_weeks11} $\Pe=\dot{\gamma} \tau$, the product of the applied shear rate $\dot{\gamma}$ and the structural relaxation time $\tau$ of the glass. When $\Pe<1$, the applied shear rate $\dot{\gamma}<\tau^{-1}$, and the relaxation is dominated by thermal fluctuations. However, when $\Pe>1$, $\dot{\gamma}>\tau^{-1}$, and relaxation is dominated by the applied shear. We define the structural relaxation time of the quiescent glass as the time that the mean square displacement of particles exceeds the particle radius. For our system, $\tau \sim 2 \times 10^4 s$, a factor of $5 \times 10^4$ larger than the Brownian diffusion time $\tau_B = 0.4 s$ in the dilute limit. We apply shear rates between $\dot{\gamma} = 1.5\times 10^{-5} s^{-1}$ and $2.2 \times 10^{-4} s^{-1}$ that correspond to Peclet numbers between $0.3$ and $4.4$, smaller and larger than one. This allows us to investigate the effect of temperature versus shear on the correlations of microscopic fluctuations.

\begin{figure}
\includegraphics[height=0.3\textwidth]{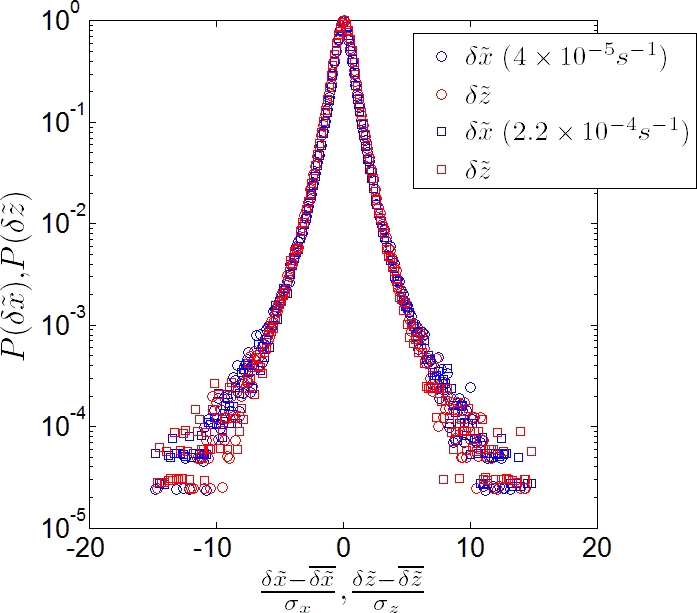}
\centering
\caption{Fluctuations of single particle displacements around the mean flow ($\Delta x$, blue symbols) and in the shear-gradient direction ($\Delta z$, red symbols). Circles and squares indicate the thermal ($\Pe=\tau\gammadot=0.8$), and shear-dominated regime ($\Pe=4.4$), respectively. All distributions overlap, demonstrating perfect isotropy.}
\label{fig:pdf_dx}
\end{figure}

\section{Experimental results and discussion}
In agreement with the idea of an effective temperature, the displacement fluctuations of the individual particles are always isotropic, as demonstrated in Fig.~\ref{fig:pdf_dx}. The figure shows the fluctuations of non-affine displacements in the flow ($x$) and the shear-gradient direction ($z$) for both $\Pe<1$ and $\Pe>1$. Clearly, the distribution of displacements in both directions are identical, and symmetric with regard to forward and backward jumps. These properties hold regardless of the Peclet number, i.e. both where thermal fluctuations determine the dynamics, and where shear is the dominant factor. We conclude that there is no shear-induced anisotropy in the fluctuations of the individual displacements, indeed favoring an "effective temperature" picture, in which the applied shear is accounted for by an isotropic effective noise.

\begin{figure}
\includegraphics[height=0.25\textwidth]{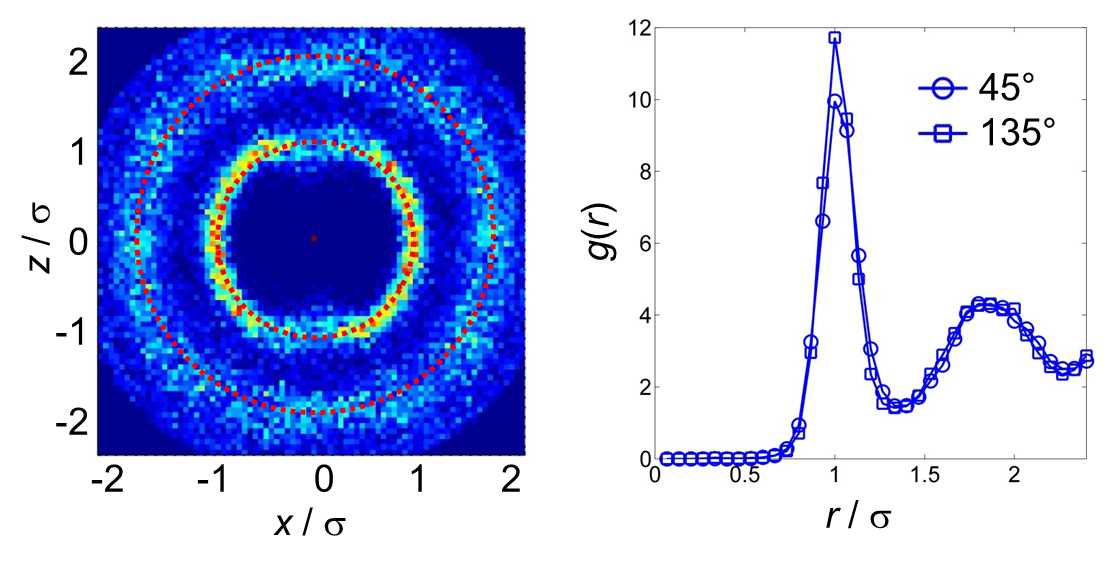}
\caption{Pair correlation function g(r) at $\Pe \gtrsim 1$, where shear dominates the relaxation. (a) Two-dimensional map resolving g(r) in the plane perpendicular to the shear axis. Dashed circles demonstrate isotropic first and second nearest neighbor shells. (b) Angle-resolved pair correlation function as a function of distance along the $45^\circ$ (axis of extension) and $135^\circ$ direction (axis of compression). Bins of width $\pi/18$ around these directions were used for averaging.}
\unitlength=1mm
\begin{picture}(0,0)
\put(0,40){(a)}
\put(45,40){(b)}
\end{picture}
\label{fig:PairCorrelationFunction}
\end{figure}
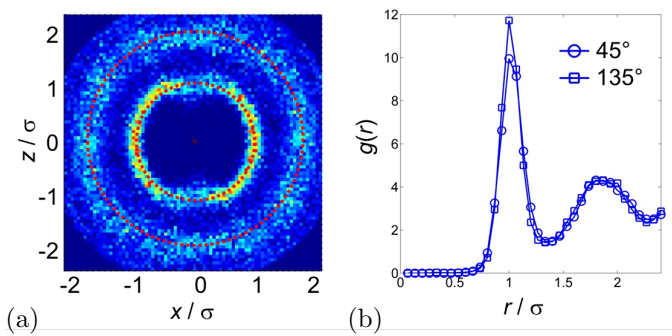

Regarding the structure of the glass, some anisotropy is expected due to the presence of a finite shear stress \cite{HansenBook_1990}. This anisotropy is, however, quite small in the range of volume fractions and shear rates of interest to this work. To demonstrate this, we determine the direction-dependent pair correlation function $g(r)$ that indicates the probability of finding two particles separated by $r$. A two-dimensional density map that resolves $g(r)$ in the plane perpendicular to the shear axis is shown in Fig.~\ref{fig:PairCorrelationFunction}a. The pair correlation function appears closely circular symmetric, indicating that the structure remains --within our experimental resolution-- essentially isotropic. The slight deviation from circular symmetry is caused by the lower instrument resolution in the vertical direction; we checked this by imaging the glass without applied shear, where we obtain a very similar $g(r)$. Nevertheless, some anisotropy is expected with respect to the two diagonal directions~\cite{HansenBook_1990}; this anisotropy is, however, very small. We resolve $g(r)$ along $45^\circ$ and $135^\circ$ to the horizontal, and plot the angle-specific $g(r)$ in Fig.~\ref{fig:PairCorrelationFunction}b. The close overlap of the two curves shows that the anisotropy is of the order of our experimental accuracy, and can barely be resolved. We note that this is in contrast to measurements performed at much higher Pe where the applied shear can have a pronounced effect on the structure, and the pair correlation becomes clearly anisotropic~\cite{Cohen2011}. At the moderate Pe investigated here, however, both the structure and the displacement fluctuations remain essentially isotropic and barely any signature of the applied shear is observable.

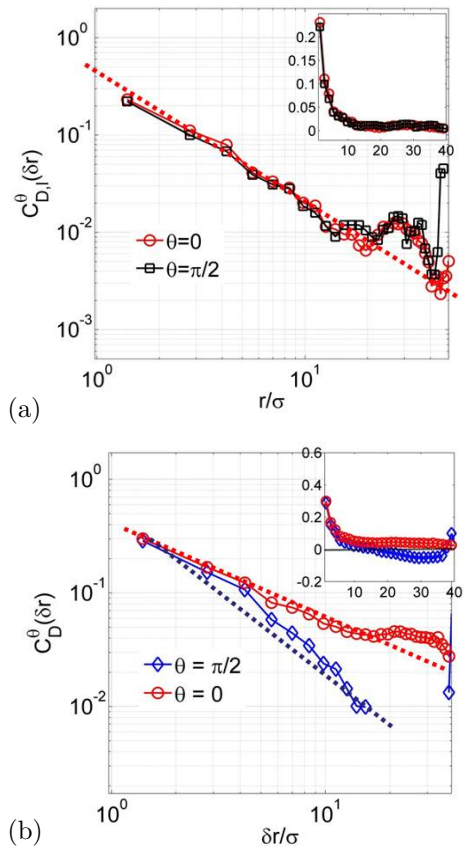
\begin{figure}
\centering
\includegraphics[height=0.3\textwidth]{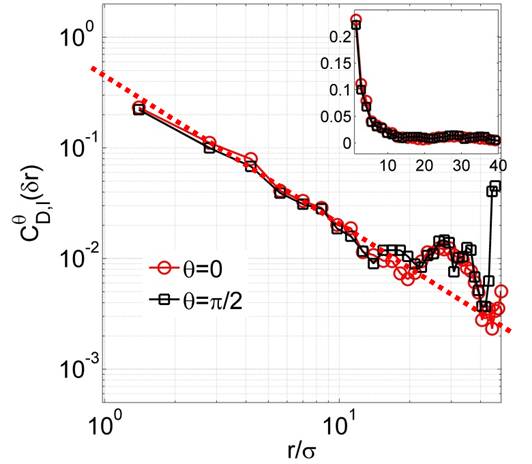}
\includegraphics[height=0.32\textwidth]{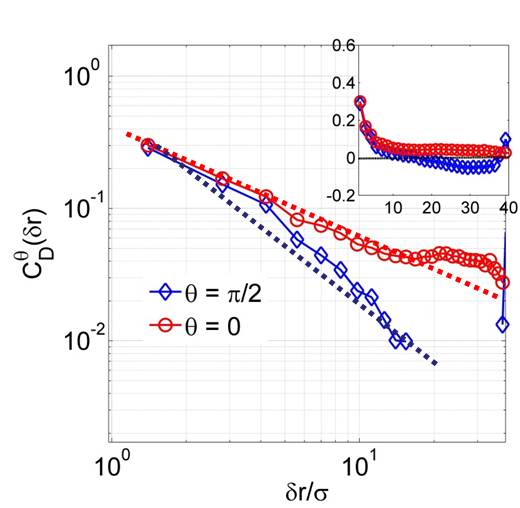}
\caption{Anisotropy of spatial correlations. Correlation function $C_{D^2}$ resolved along the horizontal (red circles) and the vertical directions (black squares and blue diamonds), in the thermal (a) and shear dominated regimes (b). Bins of width $\pi/18$ around the horizontal and vertical directions were used for averaging. $C_{D^2}$ shows isotropic algebraic decay in the thermal (a), and anisotropic decay in the shear dominated regime (b). Insets show same data on linear scale.}
\unitlength=1mm
\begin{picture}(0,0)
\put(-30,96){(a)}
\put(-30,40){(b)}
\end{picture}
\label{fig:D2-aniso}
\end{figure}

In contrast to the static correlation function and single-particle displacements, a surprisingly strong signature of the applied shear is observed in the two-point correlations of displacements. To demonstrate this, we determine the spatial correlation function \cite{Chikkadi2011}
\begin{equation}
C_{D^2}({\bf \Delta \mathbf{r}}) = \frac{ \left< D^2({\bf r + \Delta \mathbf{r}}) D^2(\mathbf{r})
\right> - \left< D^2(\mathbf{r}) \right> ^{2} } { \left< D^2(\mathbf{r})^{2}
\right> - \left< D^2(\mathbf{r}) \right> ^{2} }  ,
\label{c_r}
\end{equation}
which correlates non-affine displacements at locations separated by the vector $\Delta \mathbf{r}=(\delta x,\delta y, \delta z)$. $C_{D^{2}}$ correlates fluctuations of the plastic activity, and is analogous to correlation functions defined in critical phenomena for equilibrium fluctuations. We have recently observed that $C_{D^{2}}$ exhibits a power-law decay over the entire system size, demonstrating that plastic flow is organized in a hierarchical way \cite{Chikkadi2011}. Here, we elucidate the direction-dependence of correlations by resolving $C_{D^{2}}$ along the horizontal and vertical directions. To do so, we average $C_{D^{2}}$ in angular wedges of width $\pi/18$ and plot the angle-specific correlation function in Fig.~\ref{fig:D2-aniso}. A striking difference between the thermal and the strongly driven regime is observed. While in the thermal regime, correlations decay isotropically with the same power law $C_{D^2} \sim r^{-\alpha}$ with exponent $\alpha \sim 1.3$ (Fig.~\ref{fig:D2-aniso}a), in the stress-dominated regime, the decay becomes anisotropic: correlations decay faster in the vertical, and slower in the horizontal direction (Fig.~\ref{fig:D2-aniso}b). These results suggest that the application of shear on colloidal glasses leads to the manifestation of new correlations \cite{maloney_robbins09} with shear-rate dependent anisotropic decay.

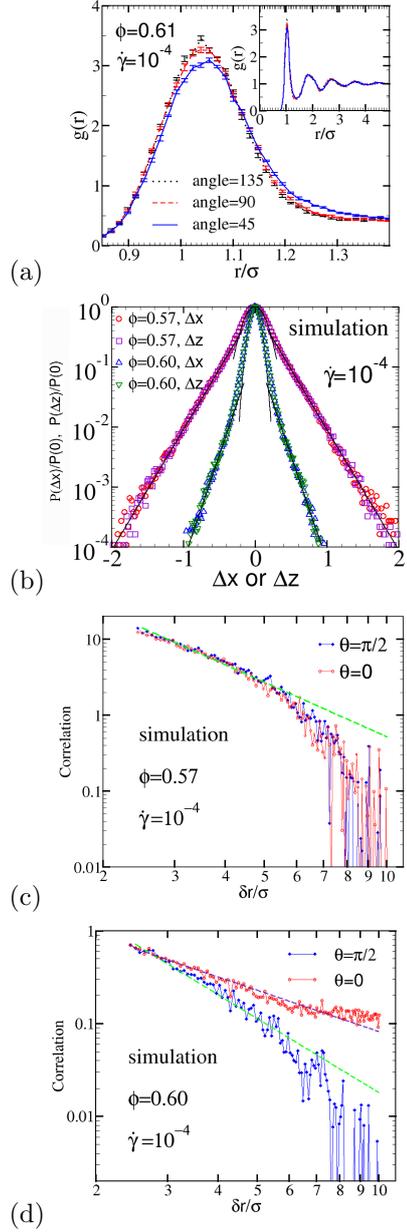
\begin{figure}
\centering
\includegraphics[height=0.22\textwidth]{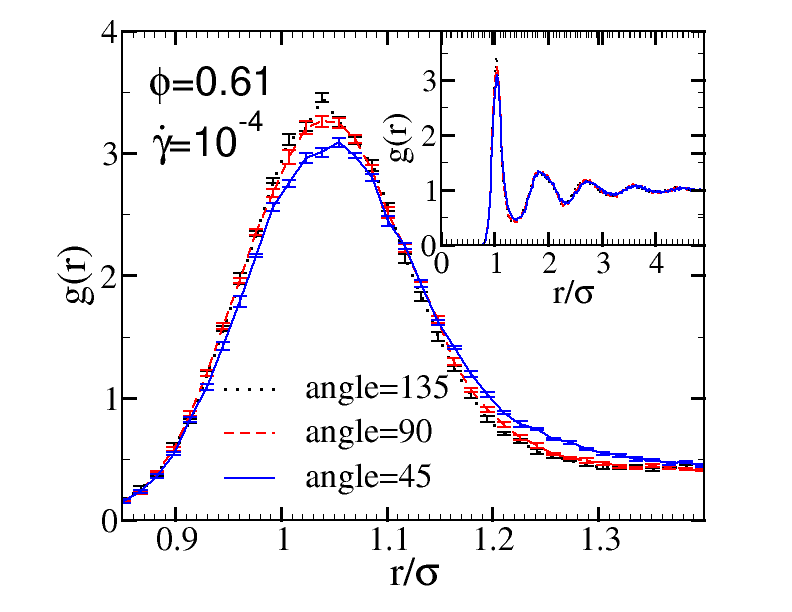}
\includegraphics[height=0.22\textwidth]{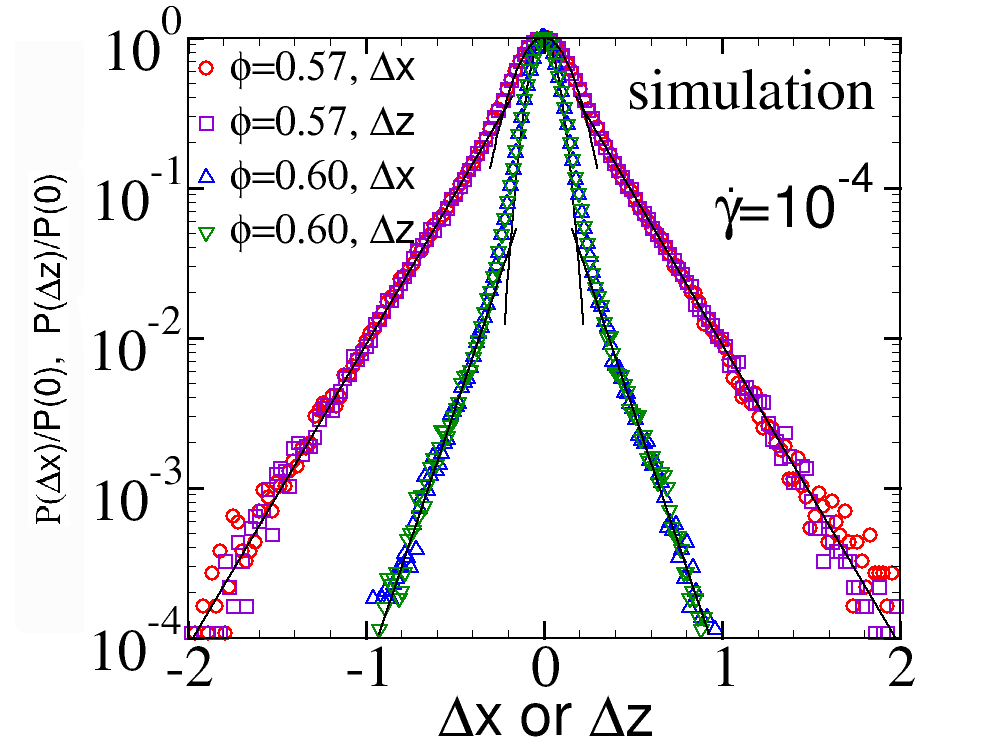}\\[1mm]
\includegraphics[height=0.23\textwidth]{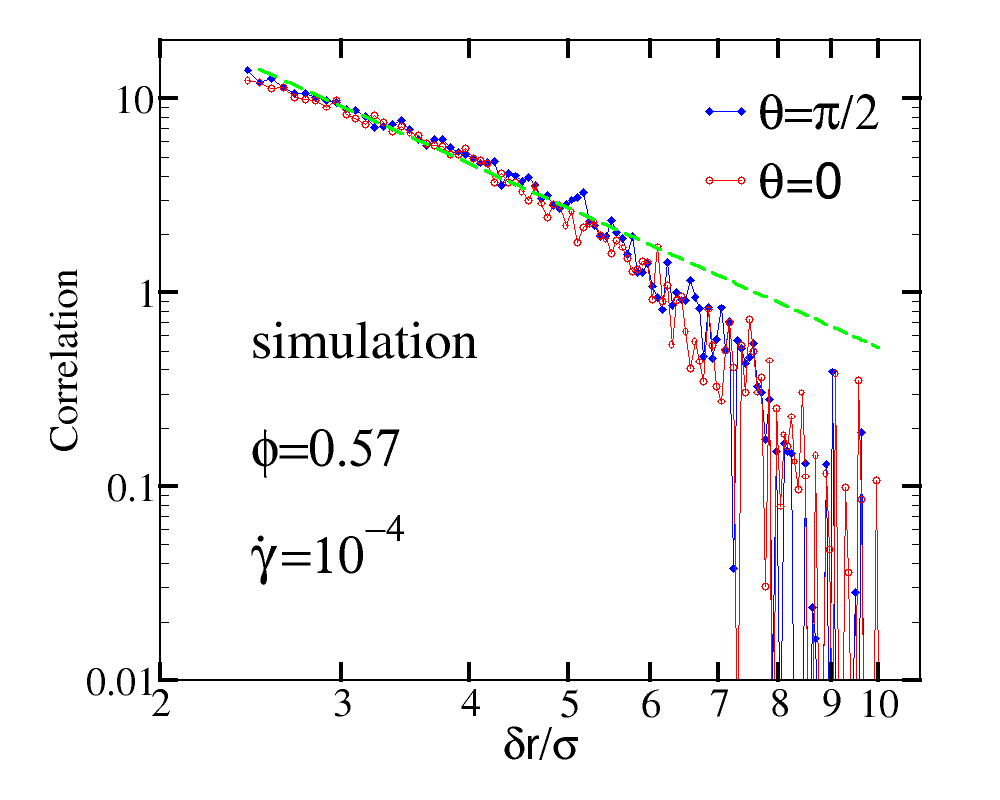}
\includegraphics[height=0.23\textwidth]{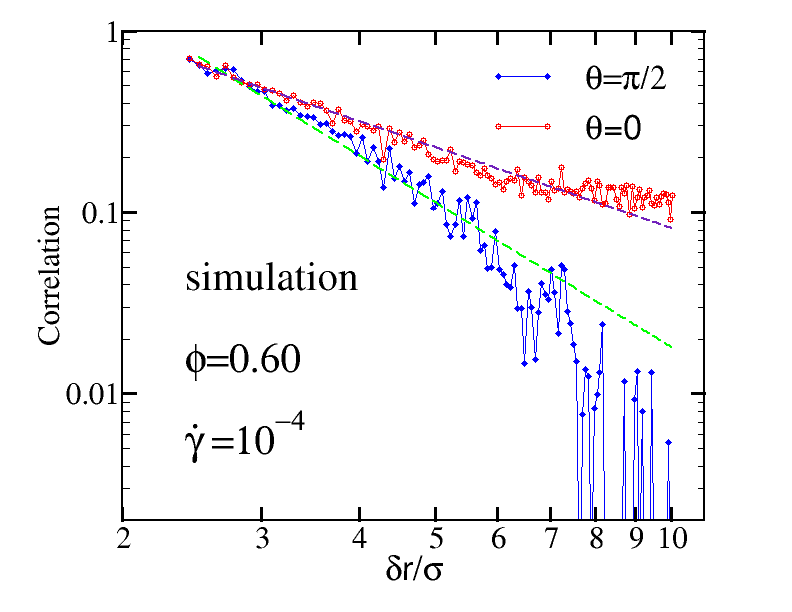}\\
\unitlength=1mm
\begin{picture}(0,0)(0,-10)
\put(-30,120){(a)}
\put(-30,79){(b)}
\put(-30,37){(c)}
\put(-30,-5){(d)}
\end{picture}
\caption{Results of wall-driven computer simulations in the thermal and shear-dominated regimes. The shear rate is $\gammadot=10^{-4}$, corresponding to Peclet numbers $\Pe=\tau\gammadot \approx 0.1$ for $\phi=0.57$ and $\Pe \ge 10$ for $\phi \ge 0.60$. (a) Pair distribution function in the shear plane (spanned by the flow and the shear-gradient vectors) along lines of $45^\circ$ (axis of extension), $90^\circ$ (flow gradient direction) and $135^\circ$ (compression axis) with respect to the flow direction. Results for the flow direction ($0^\circ$) are indistinguishable from those of the flow gradient direction. Error bars give standard deviation over 13 independent simulation runs. (b) Distribution of single particle displacements along the flow ($\Delta x$, after subtraction of the flow contribution) and flow gradient direction ($\Delta z$). The time interval for the measurement of $\Delta x$ and $\Delta z$ corresponds to 1\% strain. Dashed (solid) lines indicate gaussian (exponential) fits to small (large) displacements \cite{Chaudhuri2007}. (c) and (d) Spatial correlations of plastic activity, Eq.~(\ref{c_r}), along the directions parallel and perpendicular to the flow.}
\label{fig:simulations}
\end{figure}

\section{Simulations}
In our experiments, flow is imposed at the boundaries, and the glass flows inhomogeneously, i.e. develops shear bands in the shear-dominated regime; the question is then whether these anisotropic correlations are the result of the boundary conditions. To investigate this, we use computer simulations, where the boundary conditions can be varied more easily than in the experiment. We perform event driven MD simulations of a polydisperse (11\%) hard sphere system in three dimensions. The quiescent properties of this system have been studied extensively in \cite{Williams,Pussey2009}. The temperature is fixed at $T=1$ via velocity rescaling (in our simulations, particle mass and diameter and the Boltzmann constant are set to unity, thus setting the unit of time and energy also to 1). In the simulations, $N=30000$ particles are placed in a random configuration between two walls, separated by 30 mean particle diameters. A constant shear is applied by moving the walls with velocities $\pm \Uwall$ in $\pm x$-direction. We perform simulations in the thermal and shear-dominated regimes by keeping the shear rate constant ($\gammadot=10^{-4}$) but varying the relaxation time, $\tau$, via a change of the volume fraction from $\phi=0.57$ ($\tau \approx 10^3 \to \Pe \approx 0.1$) to a value in the glassy state, $\phi=0.60$ ($\tau > 10^5 \to \Pe > 10$). In addition to the wall-driven flow, we have also performed simulations using Lees-Edwards boundary conditions, where the top of the simulation cell is identified with the bottom displaced with a constant horizontal velocity \cite{Lees_Edwards}. While the wall-driven flow better mimics the experimental situation, the latter has the advantage of being free of wall effects, thereby revealing the bulk response of the glass \cite{VarnikPRB2006}.

The results of these simulations are shown in Fig.~\ref{fig:simulations}. Interestingly, but not unexpectedly, due to rather low stresses present in these studies, only a slight anisotropy is observed in the structure (Fig.~\ref{fig:simulations}a). Furthermore, in very good agreement with our experiments as well as previous reports \cite{VarnikSendai_08}, the fluctuations of the individual displacements exhibit essentially isotropic behavior (Fig.~\ref{fig:simulations}b). We observe a gaussian central part and a perfect exponential decay at large displacements; the universal character of these features has been the subject of recent studies (see, e.g., \cite{Chaudhuri2007} and references therein). In contrast, strong anisotropy is observed in the \textit{correlations} of displacements when the Peclet number increases. Correlations decay isotropically in the thermal regime (Fig.~\ref{fig:simulations}c), while in the shear-dominated regime, the decay is anisotropic (Fig.~\ref{fig:simulations}d). This is true for both wall-driven as well as Lees-Edwards-boundary conditions, indicating that these anisotropic correlations are rather a bulk property of the driven glass. This anisotropy seems to be a robust feature of shear-dominated flow: while the spatial decay of correlations deviate from a power law, possibly due to the different boundary conditions, the change of symmetry persists, indicating that the anisotropy is a fundamental property of shear-induced relaxation.

\section{Flow instabilities}
This anisotropic decay can have important consequences for the mechanical stability of amorphous materials. We hypothesize that it can lead to the ubiquitous shear banding instability: the longer range of correlations in the horizontal direction leads to enhanced coupling of non-affine displacements in this direction, which in turn can lead to highly correlated flow that spans the sample in the form of a shear band. At the same time, the faster decay of correlations in the vertical direction restricts the coupling of particle displacements, and constrains the shear band to a rather small thickness (see Figs.~\ref{fig:D2-aniso}b and \ref{fig:simulations}d). To test this idea, in the experiment, we investigated whether the anisotropy of correlations persist during the initial transient regime, where the flow is still uniform. Indeed, we find that already in this transient regime before any shear banding manifests, correlations are strongly anisotropic, suggesting that it is this anisotropic nature of correlations that causes the well-known shear banding instability.

\section{Conclusion}
Our experiments and computer simulations reveal a surprising stress dependence of microscopic fluctuations in glasses. Correlations of non-affine displacements are anisotropic, and exhibit unusual, stress-dependent decay. This decay is direction-dependent when shear dominates the relaxation, and becomes isotropic when thermal relaxation prevails. These results have important implications for models and theories of the plastic deformation of amorphous materials. While unifying concepts such as effective temperature \cite{FDT_barrat} and the jamming phase diagram \cite{liu_nagel98} have been quite successful in describing many aspects of the response of driven glassy systems, our results show that a more refined description should also account for the shear-induced anisotropy in spatial correlations. This shear-induced anisotropy suggests a new route for understanding the origin of flow heterogeneities in a wide range of amorphous materials. Because athermal granular materials are always dominated by the applied shear, they show always shear localization \cite{Schall_VanHecke2010, maloney_robbins09}, while thermal amorphous materials show shear instabilities only in the glassy phase but flow uniformly in the supercooled state \cite{Chikkadi2011,VarnikPRL_03,MandalPRL_12}.

\acknowledgments
P.S. acknowledges financial support by a VIDI fellowship of the Netherlands Organization for Scientific Research (NWO).

\end{document}